\documentclass[aps,prl,superscriptaddress,twocolumn,showpacs]{revtex4}
\usepackage{graphics}
\usepackage{epsfig}
\usepackage{animate}
\usepackage{color}
\usepackage{stackrel}
\usepackage{amsfonts}
\usepackage{wrapfig}
\usepackage{pstricks}
\usepackage{pst-node}
\usepackage{bm}
\usepackage{dcolumn}
\usepackage{multirow}
\usepackage{epstopdf}
\usepackage{subfigure}
\usepackage{amssymb}
\usepackage{amsmath}
\usepackage{commath}
\usepackage{graphicx,bm}
\usepackage{verbatim}

\begin{document}
\title{Theory of intervalley Coulomb interactions in monolayer transition-metal dichalcogenides}

\author{Hanan~Dery}
\altaffiliation{hanan.dery@rochester.edu}
\affiliation{Department of Electrical and Computer Engineering, University of Rochester, Rochester, New York 14627, USA}
\affiliation{Department of Physics and Astronomy, University of Rochester, Rochester, New York 14627, USA}

\begin{abstract}
Exciton optical transitions in transition-metal  dichalcogenides offer unique opportunities to study rich many-body physics. Recent experiments in monolayer WSe$_2$ and WS$_2$ have shown that while the low-temperature photoluminescence from neutral excitons and three-body complexes is suppressed in the presence of elevated electron densities or strong photoexcitation, new dominant peaks emerge in the low-energy side of the spectrum. I present a theory that elucidates the nature of these optical transitions showing the role of the intervalley Coulomb interaction. After deriving a compact dynamical form for the Coulomb potential,  I calculate the self-energy of electrons due to their interaction with this potential. For electrons in the upper valleys of the spin-split conduction band, the self energy includes a moderate redshift due to exchange, and most importantly, a correlation-induced virtual state in the band-gap.  The latter sheds light on the origin of the luminescence in monolayer WSe$_2$ and WS$_2$ in the presence of pronounced many-body interactions.
\end{abstract}
\pacs{71.45.Gm 71.10.-w  71.35.-y 78.55.-m}
\maketitle

Monolayer transition-metal dichalcogenides (ML-TMDs) have recently sparked wide interest due to their $d$-band semiconducting behavior and spin-valley coupling \cite{Radisavljevic_NatNano11,Xiao_PRL12,Zeng_NatNano12,Mak_NatNano12,Feng_NatComm12,Xu_NatPhys14,Mak_Science14,Yang_NatPhys15,Bushong_arxiv16}.  The markedly strong optical absorbance in these atomic monolayers and their compatibility with flexible substrates can enable the next generation of ultrathin photonic and optoelectronic devices \cite{Mak_PRL10,Splendiani_NanoLett10,Korn_APL11,Wang_NatNano12,Britnell_Science13,Geim_Nature13,Withers_NatMater15,Li_PRB15,Hanbicki_AIP16}. The reduced dielectric screening of Coulomb interactions and formation of tightly bound excitons in ML-TMDs \cite{Keldysh_JETP79,Cudazzo_PRB11,Cheiwchanchamnangij_PRB12,Ramasubramaniam_PRB12,Mak_NatMater13,Berkelbach_PRB13,Chernikov_PRL14,Thilagam_JAP14,Zhang_NanoLett15,Ganchev_PRL15},  allow us to study many-body interactions through the behavior of excitons in a far wider range of background plasma densities compared with typical gated semiconductor quantum wells \cite{SchmittRink_JL85,Sooryakumar_SSC85,SchmittRink_PRB86,HaugKoch_Book,Haug_SchmittRink_PQE84}.  

The motivation for this work comes from the observation of unique photoluminescence (PL) peaks that emerge in ML-WX$_2$, where X=$\{$S, Se$\}$ \cite{Jones_NatNano13,Shang_ACSNano15,You_NatPhys15,Plechinger_PPS15,Kim_ACSNano16}.  In the presence of large electron or hole densities, the PL of neutral excitons and three-body complexes decays due to screening  \cite{SchmittRink_PRB86,HaugKoch_Book,Scharf_arXiv16,Lowenau_PRL82}. However, recent experiments found that a new peak emerges in the PL of ML-WX$_2$ in the low-energy side of the spectrum when the gate-induced electron density is large \cite{Jones_NatNano13,Shang_ACSNano15}.   As shown in Fig.~\ref{fig_exp}, this peak dominates the PL of ML-WSe$_2$ at large positive gate voltages,$V_G \gtrsim  2$~V. Relative to other peaks, it shows a strong redshift when increasing the electron density \cite{Jones_NatNano13}. You \textit{et al}. have observed that the bi-exciton peak in strongly photoexcitated ML-WSe$_2$ emerges in the same spectral region \cite{You_NatPhys15}. Shang  \textit{et al}. reported of similar pattern in ML-WS$_2$ \cite{Shang_ACSNano15}. 

\begin{figure}[t]
\mbox{
\subfigure{\animategraphics[width=4.2cm,autoplay]{12}{WSe2frames}{}{}}\,
\subfigure{\animategraphics[width=4.2cm,autoplay]{12}{MoSe2frames}{}{}}}
\vspace{-0.6 cm}
\caption{Animations of the measured PL in gated ML-WSe$_2$ (left) and ML-MoSe$_2$ (right) at low temperatures \cite{footnote_thanks}. The neutral exciton is denoted by $X_0$, and the charged three-body complexes by $X_{-}$ and $X_{+}$. The many-body peak in  ML-WSe$_2$, dubbed $X^{-\prime}$, dominates the PL at elevated electron densities ($V_G \gtrsim 2$~V). I am indebted to Mitchell Jones and Xiaodong Xu for providing these results. }
\label{fig_exp}
\end{figure}

To date, there are no models that could explain why these peaks appear in the PL ML-WX$_2$ but not in that of ML-MoX$_2$ \cite{MacNeill_PRL15,Ross_NatComm13,footnote0}, or why they are not suppressed by screening as one would expect at elevated electron densities.  The only available models deal with neutral excitons or few-body complexes  \cite{Cheiwchanchamnangij_PRB12,Ramasubramaniam_PRB12,Mak_NatMater13,Berkelbach_PRB13,Chernikov_PRL14,Thilagam_JAP14,Zhang_NanoLett15,Ganchev_PRL15}. The emergence of unique peaks in the PL of strongly photoexcited  or electron-rich samples indicate the signature of many-body effects. To deal with the difference between ML-MoX$_2$ and ML-WX$_2$ and the fact that this behavior is not observed in hole-rich samples, one should also consider the subtle change in their optical transitions. The photexcitation  involves transitions from the top of the valence band to the lower (upper) valleys in the spin-split conduction band of ML- MoX$_2$ (WX$_2$)  \cite{Dery_PRB15,Zhang_PRL15}. Given that all other properties are similar, this subtlety is a key difference.

The main contribution of this Letter is the finding of particular intervalley Coulomb interactions in ML-TMDs that emerge at elevated electron densities. I show that the electron's self-energy in the upper valleys of the spin-split conduction band has a correlation-induced virtual state in the band gap, thereby affecting photoexcited excitons in ML-WX$_2$ but not in ML-MoX$_2$. As will be argued, inclusion of the intervalley Coulomb interaction provides a self-consistent explanation for the optical properties in ML-WX$_2$ when subjected to strong photoexcitation or elevated electron densities.

\begin{figure}
    \includegraphics[width=8cm]{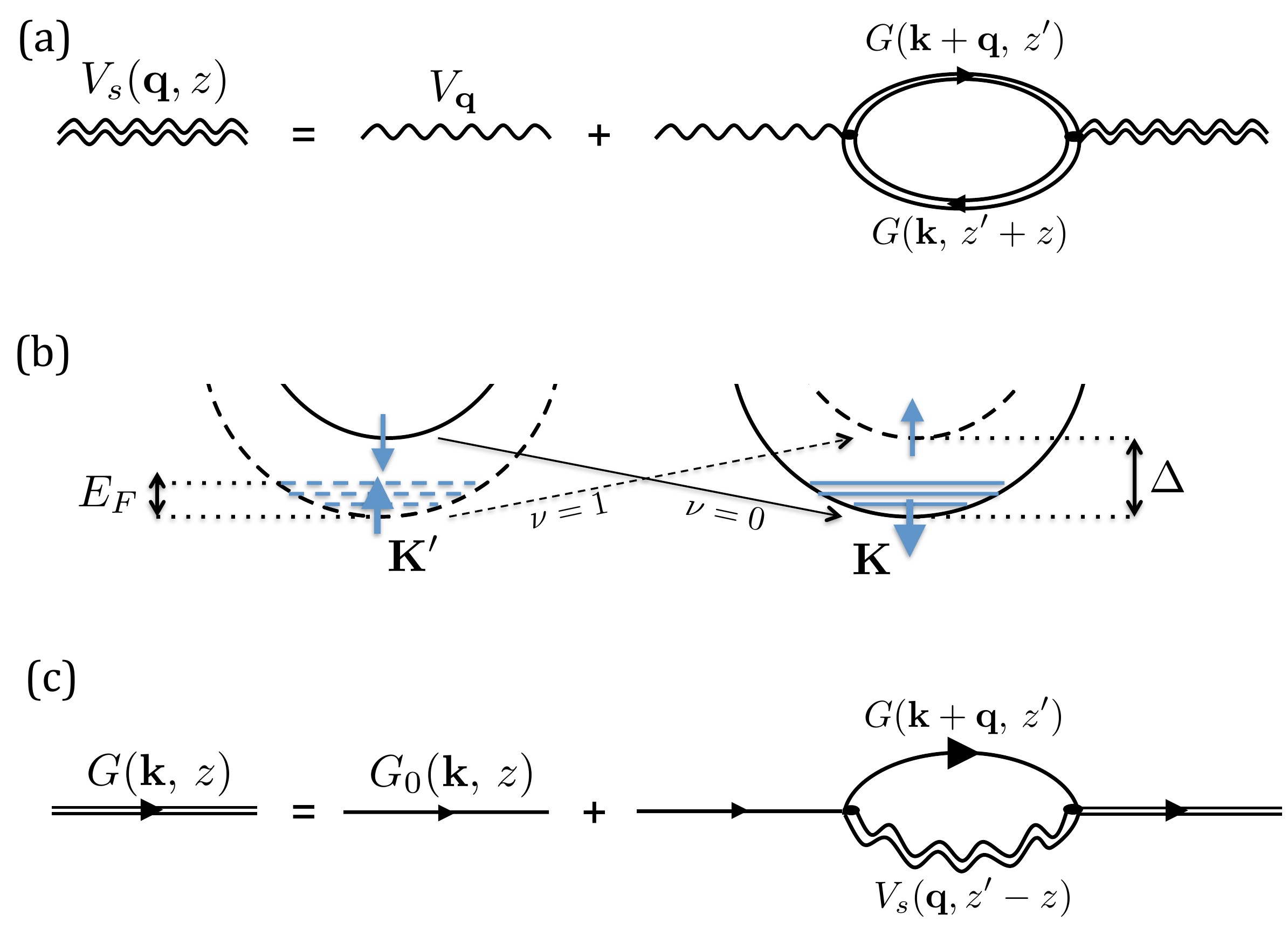}
    \caption{(a) Dyson equation for the screened Coulomb potential. The basic polarization bubble diagram includes intravalley or intervalley processes depending on whether its two propagator lines are from the same or opposite valleys. (b) The valley diagram in the conduction band of ML-TMDs. The dashed/solid lines denote valleys populated with spin-up/down electrons. The arrows represent the spin-dependent intervalley excitations. $\Delta$ and $E_F$ denote the splitting and Fermi energies. (c) The electron Green's function. The screened potential in the self-energy diagram includes both intravalley and intervalley processes.
        } \label{fig1}
\end{figure}

Before embarking on the theory, I emphasize that the intervalley Coulomb interaction in TMDs are not negligible compared with the intravalley one in the presence of elevated charge densities. Quantitatively, it can be seen by inspecting the ratio between intervalley and intravalley Coulomb interactions in the static screening limit \cite{footnote1},
\begin{equation}
\frac{V_s(\mathbf{q}\rightarrow \pm \mathbf{K}_0,\omega=0)} {V_s(\mathbf{q}\rightarrow 0,\omega=0)}  \approx  \frac{1}{K_0} \frac{2g_v}{a_B}  \,. \label{eq:ratio}
\end{equation}
$\mathbf{K}_0$ is the crystal momentum connecting the $\mathbf{K}$ and $\mathbf{K}'$ points of the hexagonal 2D Brillouin zone, where $K_0 = 4\pi/3a$  and $a\sim0.32$~nm is the in-plane sublattice constant. $2g_v/a_B$ denotes the Thomas-Fermi screening wavenumber relevant for the intravalley interaction at elevated charge densities and low temperatures, where  $g_v=2$ is the valley degeneracy and $a_B = \hbar^2 \epsilon_r / me^2$ is the effective Bohr radius \cite{Ando_RMP82}. Plugging typical values for the effective mass and dielectric constant in ML-TMDs provides $a_B \sim 0.5$~nm  ($m=0.5m_0$ and $\epsilon_r=5$). Since the Bohr radius extends over very few lattice constants, the intravalley Coulomb interaction does not overwhelm the intervalley one at elevated charge densities. As important, intervalley plasmon modes are gapped in ML-TMDs due to the spin splitting of the $K$-point. The splitting magnitude is about 20 - 30 meV in the conduction bands of ML- MoSe$_2$, WS$_2$ and WSe$_2$ \cite{Kosmider_PRB13a,Liu_PRB13,Kosmider_PRB13,Kormanyos_2DMater15}.  This attribute allows one to differentiate their signature from that of the gapless  intravalley 2D plasmons, $\omega_{q \rightarrow 0} \sim 0$ \cite{HaugKoch_Book}, in contrast with the case of graphene studied by Tudorovskiy and Mikhailov  \cite{Tudorovskiy_PRB10}.  Below I focus on intervalley plasmons and quantify their salient signatures on the self-energy of electrons.



The plasmon modes are found  from singularities in the dynamically screened Coulomb potential, $V_s(\mathbf{q},\omega) = V_{\mathbf{q}}/{\epsilon(\mathbf{q},\omega)}$, where $V_{\mathbf{q}}$ is the bare 2D Coulomb potential and $\epsilon(\mathbf{q},\omega)$  is the longitudinal dynamic dielectric function \cite{HaugKoch_Book,Pines_PR52,Pines_PR58}. Figure~\ref{fig1}(a) shows the diagram representation of the Dyson equation for $V_s(\mathbf{q},\omega)$ when using the random phase approximation and neglecting vertex corrections \cite{Haug_SchmittRink_PQE84}. Intravalley or intervalley processes are represented by the basic polarization bubble when the two propagators are from the same or opposite valleys, respectively. To account for intervalley processes ($qa \sim 4\pi/3$), the well-known Lindhard formula for $\epsilon(\mathbf{q},\omega)$  in long wavelengths ($qa \ll 1$) is recast as a matrix, where plasmon modes are found from its determinant, $| \bar{\bar{\epsilon}}(\mathbf{q},\omega) | = 0$ \cite{Tudorovskiy_PRB10}. The matrix elements represent umklapp processes due to atomic-scale local fields \cite{Adler_PR62,Wiser_PR63},
\begin{eqnarray}
\!\!&& \!\!\!\!\epsilon_{\mathbf{G},\mathbf{G}'}(\mathbf{q},z) = \delta_{\mathbf{G},\mathbf{G}'} \!-\! V_{\mathbf{q}+\mathbf{G}}\!\sum_{\mathbf{k},\nu} \!\frac{ f(\varepsilon_{\mathbf{k}})-f(\varepsilon_{\mathbf{k}+\bar{\mathbf{q}}}\!+\!\Delta)}{ (-1)^{\nu}z - \!(\Delta \!+\! \varepsilon_{\mathbf{k}\!+\!\bar{\mathbf{q}}} \!-\! \varepsilon_{\mathbf{k}}) }  \nonumber \\
&&\,\,\,\,\,\,\,\,\, \qquad \times \langle \mathbf{k}+\mathbf{q}|e^{i(\mathbf{q}+\mathbf{G}')\mathbf{r}}| \mathbf{k} \rangle \langle \mathbf{k}|e^{-i(\mathbf{q}+\mathbf{G})\mathbf{r}}| \mathbf{k}+\mathbf{q} \rangle \,.\,\,\,\,\, \label{eq:eps_G}
\end{eqnarray}
The sum has two terms, $\nu=\{0,1\}$, coming from the two spin configurations that contribute to intervalley excitations, as shown in Fig.~\ref{fig1}(b). $\mathbf{G}$ and $\mathbf{G}'$ are reciprocal lattice vectors, $z = \hbar \omega + i\delta$, $\bar{\mathbf{q}}= \mathbf{q} - \mathbf{K}_0  \,\,\,(\bar{q}a \ll 1) $, and $\Delta$ is the $K$-point spin splitting. $f(\varepsilon_{\mathbf{k}})$ and $f(\varepsilon_{\mathbf{k}+\bar{\mathbf{q}}}+\Delta)$ are Fermi-Dirac distributions in the lower and upper valleys of the spin-split conduction band, respectively. Here, $\mathbf{k}$ and $\mathbf{k}+\bar{\mathbf{q}}$ are taken with respect to their valley edge. Given the dominant contribution of the $d_{z^2}$ orbital in the conduction band \cite{Xiao_PRL12,Zhu_PRB11}, the following estimate can be used \cite{supp},
\begin{eqnarray}
\langle \mathbf{k}+\mathbf{q}|e^{i(\mathbf{q}+\mathbf{G})\mathbf{r}}| \mathbf{k} \rangle  \sim \frac{8- G^2r_d^2}{\left(4+ G^2r_d^2\right)^{3/2}}\,, \label{eq:orbital_mat_ele}
\end{eqnarray}
where $r_d$ is the effective radius of the $d_{z^2}$ orbital in the transition-metal atom. Using the facts that $r_d \sim 1\, \AA$ and $G=4\pi \ell/\sqrt{3}a$ where $\ell$ is an integer, the expression has a dominant contribution at  $\ell=0$, reducing the problem of finding plasmon modes to that of  ${\Re}e \{\epsilon_{0,0}(\mathbf{q},z)\}=0$. Assuming parabolic energy dispersion and zero net spin polarization, the intervalley plasmon energies are \cite{supp}
\begin{equation}
\hbar \omega_{\mathbf{q}} = \Delta + \varepsilon_{\bar{\mathbf{q}}} \left(1 + 3 \alpha^{-1}\right) + \tfrac{1}{3}\alpha  E_F\,, \label{eq:plasmon_modes}
\end{equation}
where $\alpha = (K_0a_B)^{-1}$ and $E_F$ is the Fermi energy. These plasmons can freely propagate when $ \varepsilon_{\bar{\mathbf{q}}}  < \tfrac{1}{9}\alpha^2 E_F$. That is, they are not Landau-damped due to electron-pair excitations in a small region of nearly perfect intervalley transitions ($\mathbf{q} \rightarrow \mathbf{K}_0$). While I have used zero-temperature analysis to derive (\ref{eq:plasmon_modes}), the result should remain valid for $\Delta > E_F \gg k_BT$, at which $\hbar \omega_{\mathbf{q}} \sim \Delta$. These are often typical conditions for all but ML-MoS$_2$ in which the spin-splitting is minute ($\Delta\sim 4$~meV \cite{Cheiwchanchamnangij_PRB13}). Finally, I use the single-plasmon pole (SPP) approximation \cite{HaugKoch_Book,Overhauser_PRB71,Rice_NC74,Zimmermann_PSSB76} to simplify the form of the intervalley  screened potential,
\begin{equation}
V_s( \mathbf{q}=\mathbf{K}_0+\bar{\mathbf{q}},\omega) \approx V_{\mathbf{K}_0}\left( 1 + \frac{f_{iv}}{(\omega + i\delta)^2 -  \omega_{\mathbf{q}}^{2}} \right)   \, . \label{eq:SPP}
\end{equation}
The residue $ f_{iv} \approx 4\alpha E_F \Delta/3\hbar^2$ is derived from the conductivity- and \textit{f}-sum rules of (\ref{eq:eps_G}), in a similar way to the case of intravalley plasmons \cite{Haug_SchmittRink_PQE84}. The pole signature scales linearly with electron density  ($f_{iv} \propto E_F \propto n$).

The compact SPP spectral form allows one to readily identify the salient features  in the electron's self energy. Performing finite-temperature Green's function analysis of the zeroth-order diagram, shown in Fig.~\ref{fig1}(c), the self-energy due to the intervalley interaction follows
\begin{equation}
\!\!\!\! \Sigma(\mathbf{k},z) \!=\! - \frac{ 3V_{\mathbf{K}_0}}{\beta} \!\sum_{\bar{\mathbf{q}},z'} \!\left[ G_0(\bar{\mathbf{q}},z') + \frac{\hbar^2f_{iv}G_0(\bar{\mathbf{q}},z')}{(z-z')^2 - (\hbar\omega_{\mathbf{q}})^2 }\!\right]\! , \label{eq:self}
\end{equation}
where $\beta^{-1}=k_BT$ and $G_0$ is the free-electron Green's function. The first and second terms in the sum correspond to exchange and correlation, respectively. To quantify their contributions, the sum over Matsubara frequencies ($z'$) is replaced with contour integration  \cite{Haug_SchmittRink_PQE84}. The exchange self energies due to the intervalley interaction in the lower and upper valleys are, respectively,
\begin{eqnarray}
\Sigma_x^l( \mathbf{k}) = 0  \,\,\,\,, \qquad \Sigma_x^u( \mathbf{k}) \approx - \frac{2\pi n e^2}{g_v \epsilon_r K_0}  \, . \label{eq:exchange}
\end{eqnarray}
The energy in a lower valley is not renormalized due to the vanishing electron population in the opposite upper valley, $f(\varepsilon_{\mathbf{k}+\bar{\mathbf{q}}}\!+\!\Delta) \rightarrow 0$. Conversely, the nonzero energy renormalization in an upper valley comes from electron filling in the opposite lower valley. Overall, the exchange effect is small and largely wavevector-independent resulting in a rigid redshift of the upper valleys by about  1~meV per $n$=10$^{12}$~cm$^{-2}$ in the lower valleys.

\begin{figure}
    \includegraphics[width=9cm]{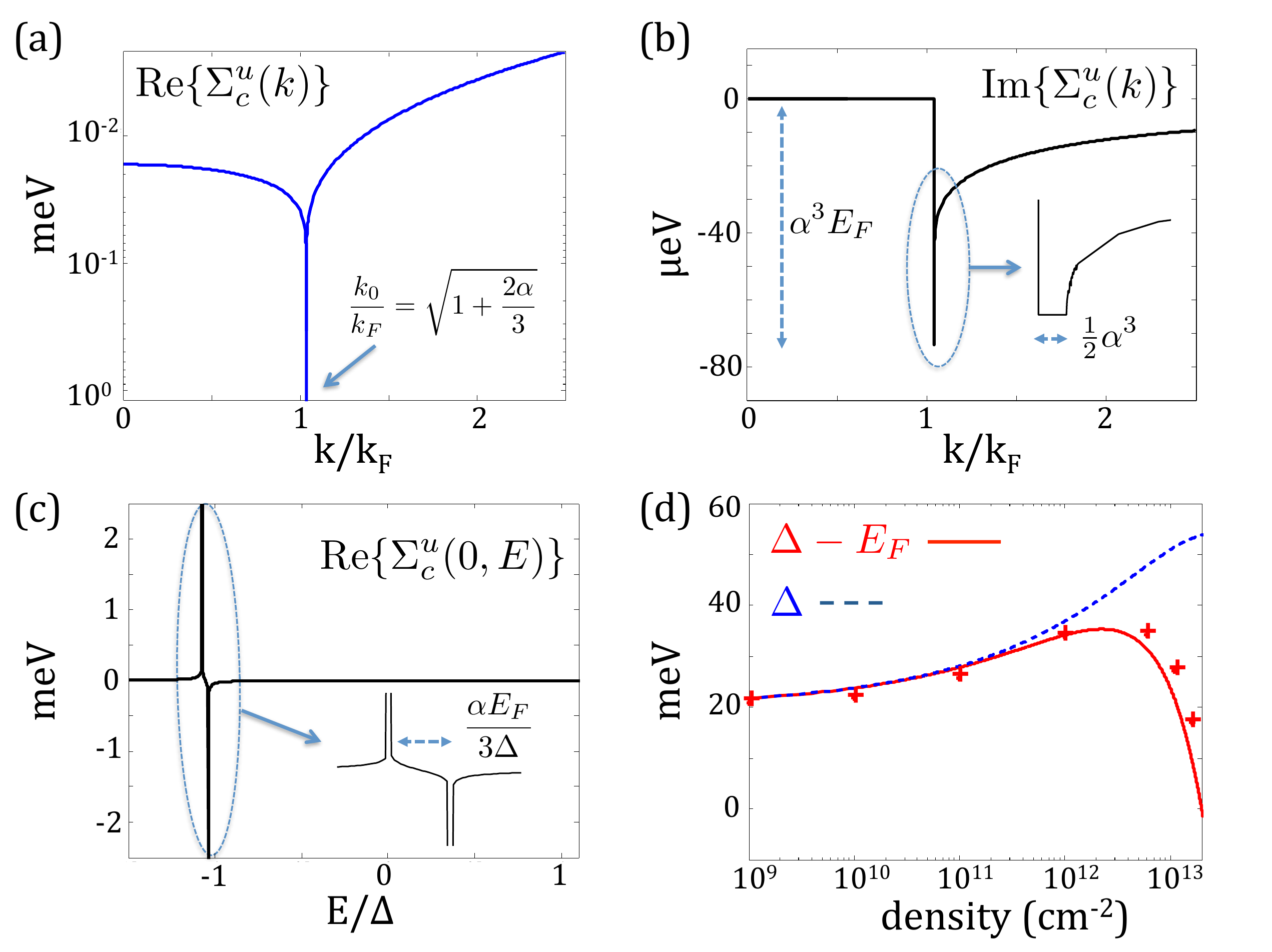}
    \caption{(a)-(c): Effects of intervalley Coulomb correlations on the self energy of electrons in the upper valleys for $n$=8$\times$10$^{12}$~cm$^{-2}$  in the lower valleys and $\alpha$=0.115. (a)/(b) Real/Imaginary parts of the renormalization energy, $\Sigma_c^u(k) \equiv \Sigma_c^u(k,\Delta+\varepsilon_{\mathbf{k}}-\mu + \i\delta)$. The real part is singular slightly above $k_F$, at which the imaginary part is finite reaching $-\alpha^3E_F$ over a small region as shown in the inset. (c) $\Sigma_c^u(\mathbf{k}=0,E)$, where $E=0$ denotes the edge of the lower valleys. The double-resonance feature in the band gap, spaced by $\sim\alpha E_F/3$, is located about $\Delta$ ($2\Delta$) below the edge of the lower (upper) valleys. (d) Renormalized conduction-band splitting ($\Delta$) and of $\Delta-E_F$ as a function of electron density. 
        } \label{fig2}
\end{figure}

Coulomb correlations in the self energy are calculated by repeating the analysis for the second term in~(\ref{eq:self}). The correlation term in the lower  valleys follows
\begin{eqnarray}
\Sigma_c^l( \mathbf{k},z-\mu) = \frac{ \alpha^2 E_F}{\pi}\frac{\hbar^2}{m}  \int \! d^2\bar{q}  \frac{ g(-\omega_{\mathbf{k}-{\mathbf{q}}})  }{z- \varepsilon_{\bar{\mathbf{q}}}  -  \Delta -   \hbar  \omega_{\mathbf{k}-{\mathbf{q}}} }, \label{eq:correlation_low}
\end{eqnarray}
and in the upper valleys,
\begin{eqnarray}
\Sigma_c^u( \mathbf{k},z-\mu) &=&  \Sigma_c^l( \mathbf{k},z+\Delta -\mu)  \label{eq:correlation_upper} \\
&+& \!\!\!\!  \frac{ \alpha^2 E_F}{\pi}\frac{\hbar^2}{m} \! \int \! d^2\bar{q}  \frac{2 \hbar \omega_{\mathbf{k}-{\mathbf{q}}}  f(\varepsilon_{\bar{\mathbf{q}}})  }{(z- \varepsilon_{\bar{\mathbf{q}}})^2 - (\hbar \omega_{\mathbf{k}-{\mathbf{q}}})^2 } \,\,, \, \nonumber
\end{eqnarray}
where $\mu$ is the chemical potential and $g(-\omega_{\bar{q}}) \rightarrow 1$ is the Bose-Einstein distribution ($\Delta \gg k_BT$). $\Sigma_c^u$ is affected by plasmon emission, denoted by the first line in (\ref{eq:correlation_upper}), and electron filling in the opposite lower valley, denoted by the second line in (\ref{eq:correlation_upper}). In the following, the integration is limited to the region of free plasmon propagation, $|\mathbf{k}-\bar{\mathbf{q}}| < \alpha k_F/3$ where $k_F$ is the Fermi wavevector. Using the fact that $\alpha^2/9 \ll 1$, the principal value of (\ref{eq:correlation_low}) is singular at $z = \varepsilon_{\mathbf{k}} + 2\Delta -\mu  + i\delta$. At these energies, plasmon emission to the opposite valley is enabled.  The renormalization of electron energies in the lower valley, resolved from  $z = \varepsilon_{\mathbf{k}}$ in (\ref{eq:correlation_low}), is negligible. On the other hand, the energy renormalization in the upper valleys, resolved from  $z = \Delta+\varepsilon_{\mathbf{k}}$ in (\ref{eq:correlation_upper}), is non-negligible in the immediate vicinity of $ k_0^2 = (1+2\alpha/3)k_F^2 $ at which the integral is singular, as shown in Fig.~\ref{fig2}(a). The imaginary part, shown in Fig.~\ref{fig2}(b), corresponds to the rate of plasmon emission. It vanishes for $k < k_0$ and decays to zero from about $-\alpha^3 E_F$ for $k > k_0$. 
The second effect, coming from the second line in (\ref{eq:correlation_upper}), yields additional logarithmic singularity. As shown in Fig.~\ref{fig2}(c), a double resonance spectral feature emerges for states in the bottom of the upper valleys at about $z \sim \varepsilon_{\mathbf{k}} - \Delta$ (i.e., within the band-gap). It can be understood as plasmon mediated virtual transition due to electron filling in the opposite lower valley. Note that the singularities in $\Sigma_c^u$, shown in Figs. \ref{fig2}(a) and (c), are robust; they withstand stringent integration over the small region of free plasmon propagation.

Before discussing the implication of these results, a point on the role of the intravalley Coulomb interaction is in place. Most relevant is the dependence of the spin splitting on population  of the lower valleys,
\begin{eqnarray}
\Delta = \Delta_0 + \Sigma_x^u - \Sigma_{0}\,, \label{eq:delta_comp}
 \end{eqnarray}
where $\Delta_{0}= \Delta_{so} + \Delta_{e-h}$ is the splitting induced by spin-orbit coupling \cite{Xiao_PRL12,Zhu_PRB11,Song_PRL13} and electron-hole exchange \cite{Maialle_PRB93,Yu_NatComm14,Echeverry_arXiv16}. $\Sigma_x^u $ is the aforementioned small redshift of the upper valleys due to intervalley-induced exchange. $\Sigma_{0}$ is the rigid redshift of the populated lower valleys in ML-TMDs due to the intravalley Coulomb interaction \cite{Haug_SchmittRink_PQE84,footnote_excited_state}. This redshift is calculated by repeating the above procedures with using the appropriate SPP form in the long range limit \cite{HaugKoch_Book}: replacing the residue $f_{iv} \rightarrow \omega_{\text{pl},q}^2$ and pole energy $\hbar^2 \omega_{\mathbf{q}}^2 \rightarrow \hbar^2\omega_{\text{pl},q}^2(1 + q/\kappa) + C\varepsilon_{q}^2$, where $\omega_{\text{pl},q} \simeq \sqrt{2\pi e^2 n q /m\epsilon_r}$ is the plasma frequency, $\kappa$ is the Thomas-Fermi screening wavenumber, and $C\varepsilon_{q}^2$ reflects the role of pair excitations ($C \sim 1$) \cite{Haug_SchmittRink_PQE84,Overhauser_PRB71}. Schmidt-Rink and Ell found that $\Sigma_{0} \simeq -C_1E_0(a_B^2n)^{1/3}$ \cite{SchmittRink_JL85},  where $E_0$ is the effective Rydberg energy and $C_1$ is a constant that depends on the integration cutoff \cite{supp}. The dashed line in Fig.~\ref{fig2}(d) shows this redshift when using $\Delta_0 = 20$~meV and $C_1$=0.6. Another important parameter is $\Delta - E_F$, which denotes the energy spacing between the conduction edge of the upper valleys and the Fermi level in the lower valleys (see Fig.~\ref{fig1}(b)). The symbols denote numerical results \cite{supp}. As shown by the solid line in  Fig.~\ref{fig2}(d), $\Delta - E_F$  is mostly governed by $\Delta$ due to the redshift of the lower valleys at low densities and by $E_F$ at large densities (i.e., competition between $n^{1/3}$ and $n$). Population of the upper valleys, $\Delta < E_F $, starts at $n \sim 10 ^{13}$~cm$^{-2}$. 

Putting these pieces together, the intervalley Coulomb interaction offers a self consistent explanation for recent experiments in ML-WX$_2$ \cite{Jones_NatNano13,Shang_ACSNano15}. In these materials, bright direct excitons are formed from states in the upper valleys \cite{Dery_PRB15,Zhang_PRL15}. At elevated electron densities, the self energies of these states include a resonance in the band-gap due to the intervalley Coulomb correlations, as shown in Fig.~\ref{fig2}(c). Optical transitions in ML-WX$_2$ can therefore be mediated by shaking up the Fermi sea via creation of intervalley plasma excitations. Importantly, the attraction to holes is not weakened by screening from the background electrons due to the shortwave and fast oscillation of these excitations ($K_0 > \kappa$ and $\Delta/\hbar > \omega_{pl,q}$). Their  signature in the optical spectrum can be resolved due to the spin splitting in the  conduction band, contrary to the gapless intravalley plasma excitations in quantum wells \cite{Sooryakumar_SSC85}. Roughly, the emerged PL peak should appear $\sim 2\Delta$ below that of neutral excitons, formed by the free states in the upper valleys. Since the lower valleys redshifts much strongly than the upper ones ($\Sigma_0$ vs $\Sigma_x^u$ in (\ref{eq:delta_comp})), the emerged peak redshifts with increasing electron density. Such behavior is not observed in ML-MoX$_2$ since its excitons are formed from states in the lower valleys. In this case, only the intravalley plasma excitations can affect the optical properties.

The same physical picture is offered to explain the emerged PL peaks in strongly photoexcited ML-WX$_2$ \cite{Shang_ACSNano15,You_NatPhys15,Plechinger_PPS15,Kim_ACSNano16}. Here, intervalley excitations can take place between the electron components of direct and indirect bright excitons while the holes are `spectators'. In ML-TMDs, the exciton band structure is comprised of a direct branch in the zone center and an indirect branch in the zone edge \cite{Dery_PRB15,Song_PRL13,Wu_PRB15}. The direct exciton branch is associated with electrons and holes from the same region in the Brillouin zone (e.g., both from the $K$-point valley), and the indirect exciton branch with pairs from opposite valleys.  In addition, these branches are split to optically active (bright) and inactive (dark) excitons depending on their spin configuration. The direct-bright energy branch is located below that of the direct-dark in ML-MoX$_2$, and above it in ML-WX$_2$  \cite{Dery_PRB15}. The indirect-bright and indirect-dark branches have the opposite order. As a result, photogenerated excitons in the direct-bright branch can undergo energy relaxation to the indirect-bright branch only in ML-WX$_2$.  Intervalley excitations between direct and indirect bright excitons are therefore viable in  strongly photo-excited ML-WX$_2$, supporting the emergence of the many-body resonance peak \cite{footnote2}. Clearly, the spin-split conduction band and the valley degree of freedom renders the physics intriguing compared with that of conventional biexciton luminescence. 

In conclusion,  I have presented a theory for the intervalley Coulomb interaction in monolayer transition-metal dichalogenides, finding their energy dispersion and effect on the self-energy of electrons. The resulting shortwave plasmons are gapped due to the spin-splitting of the energy bands in the $K$ and $K'$ points of the Brillouin zone, and their energy increases due to the redshift of the lower valleys induced by the intravalley (long-range) Coulomb interaction. Importantly, states in the upper valleys are affected by intervalley Coulomb correlations through the emergence of a resonance in the band-gap at elevated electron (or exciton) densities. This result is central to the difference in the luminescence properties of tungsten-based and molybdenum-based compounds. In addition to providing a self-consistent picture to explain experimental findings, the presented theory should lead to further investigations. These include: (1) studying excitons using the Bethe-Salpeter equation in which the electron Green's function is dressed by the intervalley Coulomb interactions. (2) Studying spin selective intervalley interactions by polarizing the background electrons via optical valley orientation or by application of a large magnetic field in the regime where the Zeeman energy is larger than the thermal energy. (3) Studying plasmon-phonon coupling, making use of the fact that $\Delta$ increases with population of the lower valleys. This increase can lead to resonance crossing with zone-edge phonons, providing alternative explanation for the observed splitting of the charged exciton in tungsten-based compounds \cite{Jones_NatPhys16}, which so far was attributed to electron-hole exchange.

\acknowledgments{I am grateful to Xiaodong Xu, Jie Shan, and Kin Fai Mak  for many useful discussions and for sharing valuable experimental results prior to their publication. I am grateful to Refik Kortan for encouraging me to pursue this topic. This work was supported by the Department of Energy under Contract No. DE-SC0014349, National Science Foundation under Contract No. DMR-1503601, and the Defense Threat Reduction Agency under Contract No. HDTRA1-13-1-0013.}


\begin{widetext}

\noindent {\bf Derivation of Eq.~(3) in the main text.}  The dominant contribution of the $d_{z^2}$ orbital leads to
\begin{eqnarray}
\langle \mathbf{k}+\mathbf{q}|e^{i(\mathbf{q}+\mathbf{G})\mathbf{r}}| \mathbf{k} \rangle  \sim \frac{5}{16\pi} \int_{0}^{\infty}  dr r^2 e^{-2r/r_d} \int_{0}^{\pi} d\theta \sin \theta (3\cos\theta^2-1)^2 \int_{0}^{2\pi} d\phi e^{iGr\cos\phi}=  \frac{8- G^2r_d^2}{\left(4+ G^2r_d^2\right)^{3/2}}\,, \label{eq:orbital_mat_ele}
\end{eqnarray}

\noindent {\bf Derivation of intervalley plasmon dispersion [Eq.~(4) in the main text].}  As mentioned in the discussion of the main text after Eq. (3), the finding of plasmon dispersion reduces to that of   ${\Re}e \{\epsilon_{0,0}(\mathbf{q},z)\}=0$. Figure~\ref{fig_bz} shows the six possible intervalley transitions from the $K'$ valley to the $K$ valley for $\mathbf{K_0} = K_0 \hat{x}$.  Only two of the six transitions, $i' \rightarrow i$, are limited to the first Brillouin zone ($i=\{1,6\}$), while the other four are umklapp processes (i.e., $G\neq0$ for $i=\{2-5\}$). Thus, in transforming the sum in the dielectric function (Eq.~(2) of the main text), one should factor the sum by $1/3$.  The reduced plasmon equation is then written as,
\begin{eqnarray} \label{eq:eps00_A}
\epsilon_{0,0}(\bar{\mathbf{q}},\omega)  = 0 = 1 -  \tfrac{1}{3} V_{K_0} \sum_{\mathbf{k}}  f_{\mathbf{k}} \left[ \frac{1}{\hbar\omega_{\bar{\mathbf{q}}} - (\Delta + \varepsilon_{\mathbf{k}+\bar{\mathbf{q}}} - \varepsilon_{\mathbf{k}})} -   \frac{1}{\hbar\omega_{\bar{\mathbf{q}}} + (\Delta + \varepsilon_{\mathbf{k}+\bar{\mathbf{q}}} - \varepsilon_{\mathbf{k}})}\right]
\end{eqnarray}
where $V_{K_0} =  2\pi e^2 / A \epsilon_r K_0$. The first and second term in square brackets imply that if $\hbar\omega_{\bar{\mathbf{q}}}$ is a plasmon mode then $-\hbar\omega_{\bar{\mathbf{q}}}$ should be one as well. Given that the plasmon modes should appear close to $\Delta$, only one of the two terms in the sum will provide a dominant contribution. I continue with the first term and assume zero temperature and no net spin polarization ($f_{\mathbf{k}} = 1$ for $k<k_F$).  Considering parabolic dispersion, I  transform the sum into the following implicit equation,
\begin{eqnarray} \label{eq:eps00_B}
1  =  \frac{\alpha}{3\pi} \int_0^{k_F} dk\, k \int_0^{2\pi} d\theta  \frac{1}{k_{\bar{q}}^2 + 2k\bar{q}\cos\theta} = \frac{\alpha}{6\bar{q}^2} \left[   k_{\bar{q}}^2  -  \sqrt{k_{\bar{q}}^4 - 4 k_F^2 \bar{q}^2}  \right],\,\,\, \alpha \equiv \frac{ 3a me^2}{4\pi \hbar^2 \epsilon_r }\,,\,\,\,\,   \hbar\omega_{\bar{\mathbf{q}}}= \frac{\hbar^2 k_{\bar{q}}^2}{2m} + \varepsilon_{\bar{\mathbf{q}}}  + \Delta.
\end{eqnarray}
The solution is trivial
\begin{equation}
\hbar \omega_{\bar{\mathbf{q}}} = \Delta + \varepsilon_{\bar{\mathbf{q}}} \left(1 + 3 \alpha^{-1}\right) + \tfrac{1}{3}\alpha  E_F\,\,\,\,\,, \qquad  \bar{q} < \frac{\alpha}{3}k_F \label{eq:eps00_D}
\end{equation}
The restriction of $\bar{q}$ to very small values in which intervalley plasmons propagate without damping, reinforces the use of only the first term in (\ref{eq:eps00_A}) for positive energies (i.e., the correction by the second term is negligible).

\begin{figure}
    \includegraphics[width=6.5cm]{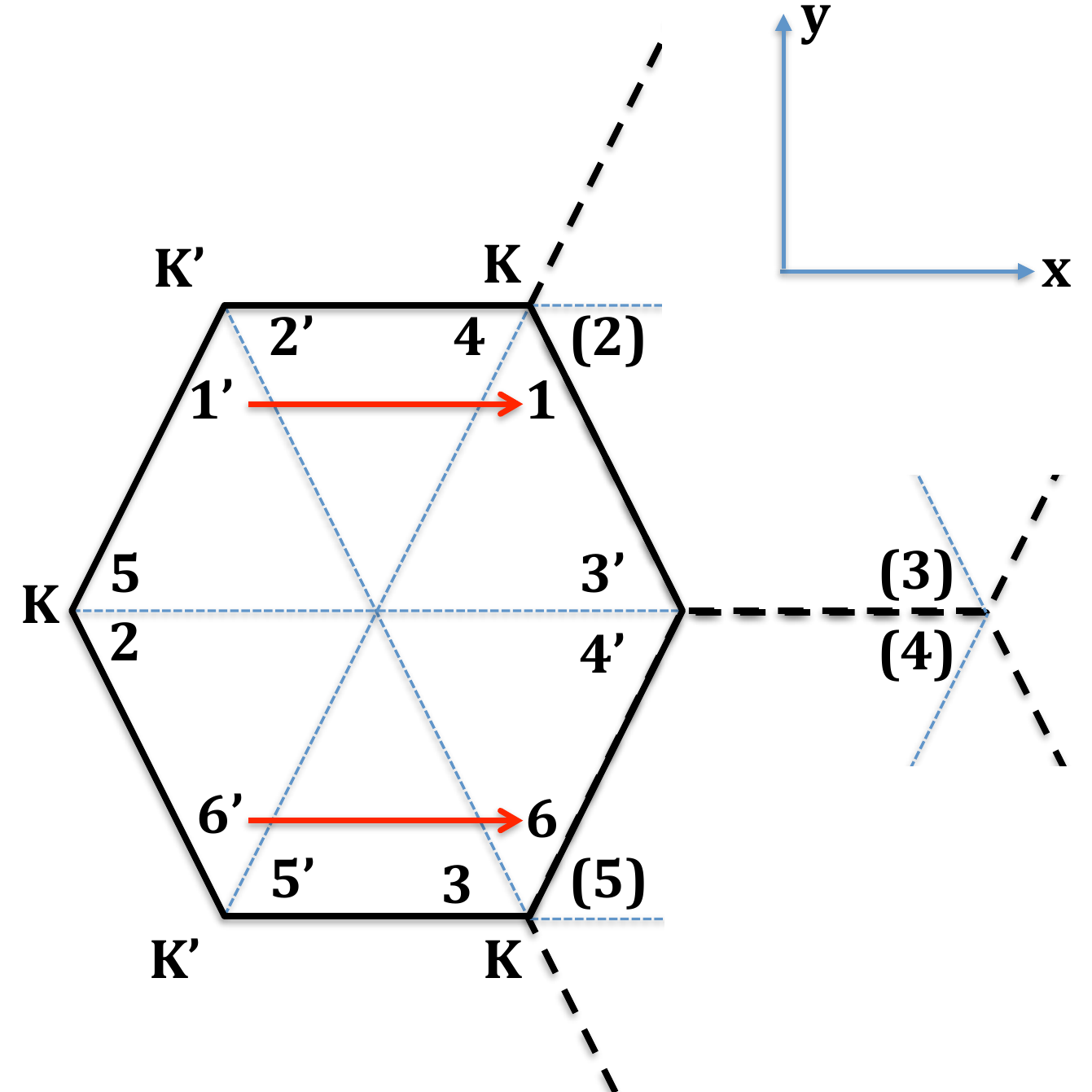}
    \caption{Intervalley scattering for $\mathbf{K_0} = K_0 \hat{x}$.
        } \label{fig_bz}
\end{figure}


\vspace{3mm}

\noindent {\bf Derivation of the SPP form [Eq.~(5) in the main text].}
In replacing the cumbersome expression for the longitudinal dynamic dielectric function with a simplified single-plasmon pole expression, I have satisfied the conductivity and $f$-sum rules [see, e.g., H. Haug and S. Schmitt-Rink, Prog. Quant. Electr. \textbf{9}, 3 (1984)] . For the intervalley case, the simplified form of the dielectric function reads
\begin{equation}
\frac{V_{\mathbf{K}_0}}{\tilde{\epsilon}(\bar{q},\omega)}  \approx   V_{\mathbf{K}_0}\left( 1 + \frac{f_{iv}}{(\omega + i\delta)^2 - \omega_{\bar{q}}^2} \right)   \,\,\,\,, \qquad \tilde{\epsilon}(\bar{q},\omega) = 1 - \frac{f_{iv}}{(\omega + i\delta)^2 - (\omega_{\bar{q}}^2 - f_{iv})} . \label{eq:SPP_A}
\end{equation}
To satisfy the conductivity sum rule we should compare the SPP  and random-phase approximation  forms,
\begin{equation}
\int_0^{\infty} d\omega  \omega \text{Im} \{  \epsilon_{0,0}(\bar{\mathbf{q}},\omega)  \}  =   \int_0^{\infty} d\omega \omega  \text{Im} \{   \tilde{ \epsilon}(\bar{q},\omega)  \}. \label{eq:SPP_B}
\end{equation}
Substituting (\ref{eq:eps00_A}) and (\ref{eq:SPP_A}) in left and right hand sides, respectively, and using Dirac identity one gets,
\begin{equation}
\frac{V_{K_0}}{3\hbar^2} \sum_{\mathbf{k}} (\Delta + \varepsilon_{\mathbf{k}+ \bar{\mathbf{q}}} - \varepsilon_{\mathbf{k}}) f_{\mathbf{k}}  =   \frac{f_{iv}}{2} . \label{eq:SPP_C}
\end{equation}
Assuming parabolic dispersion, the residue is readily resolved,
\begin{equation}
 f_{iv} = \frac{4\alpha E_F (\Delta+\varepsilon_{\bar{\mathbf{q}}})}{3\hbar^2}  \approx  \frac{4\alpha E_F \Delta}{3\hbar^2}   \label{eq:SPP_C}
\end{equation}

\vspace{3mm}

\noindent {\bf Calculation of $\Sigma_0$ (Eq.~(10) and Fig.~2(d) of the main text). }
The redshift of the lower spin-split conduction band is calculated using Eqs.~(6)-(8) in S. Schmitt-Rink and C. Ell, J. Lumin. \textbf{30}, 585 (1985). This calculation seems to overestimate the redshift compared with the one measured from the shift of the spectral peaks in the PL of ML-TMDs. The reason for this discrepancy stems from the treatment of the ultraviolet singularity (i.e., shortwave limit). Specifically, the correlation integral for long-range Coulomb excitations (intravalley) converges due to the phenomenological insertion of the particle-hole excitation to deal with short wavelengths [see Eq. (8) in J. Lumin. \textbf{30}, 585 (1985)] . This insertion is not as problematic at relatively low densities applicable to typical semiconductor quantum wells.  Schmitt-Rink and Ell found that the redshift ($\Sigma_0$ in Eq. (10) of the main text) behaves close to $\Sigma_{0} \simeq -C_1E_0(a_B^2n)^{1/3}$,  where $E_0$ is the effective Rydberg energy and $C_1 = 3.1$ for exciton plasma when the electron and hole masses are equal (or $C_1=1.55$ for a single plasma component relevant for our discussion).

To better match the experimental result in ML-TMDs, I chose an alternative approach in which the integration cutoff is limited to 2$k_F$ (or, equivalently, 4$E_F$), due to the change in the behavior of screening at larger values [see discussion in Sec. II.C in T. Ando, A. B. Fowler, and F. Stern, Rev. Mod. Phys. \textbf{54}, 437 (1982)].  This choice yields a moderate redshift due to intravalley exchange and correlation, as shown in Fig.~2(d) of the main text. Similarly, this choice amounts to changing the value of $C_1$ from 1.55 to 0.6, as shown by the dashed line in Fig.~2(d).

\vspace{3mm}

\noindent {\bf Hole-rich systems. } If the upper spin-split valence band is populated with holes, then intervalley plasmons should also play a role for type B excitons (formed by transitions from the lower spin-split valence band). However, their detection could be elusive since the transfer rate of excitons from type B to type A is ultrafast (much faster than the exciton recombination lifetime), leaving no time for the intervalley Coulomb excitation to dress the excitons. In this view, the smallness of $\Delta$ in the conduction band renders intervalley Coulomb excitations relevant in electron-rich ML-TMDs. For example, if $\Delta-E_F$ is smaller than the  energy of zone-edge phonons,  a bottleneck in the energy relaxation of excitons is created, leaving enough time for intevalley Coulomb excitations to dress the excitons. This fact can be recognized from the comparable PL intensities in MoX$_2$ and WX$_2$ \cite{Wang_PRB14,Wang_APL15,Robert_arXiv16}, implying that the recombination lifetime is comparable or shorter than the intervalley energy relaxation (otherwise the PL intensity in WX$_2$ should be much weaker). Another support is provided by the opposite-sign contributions of the spin-orbit coupling and electron-hole exchange, which overall diminish the magnitude of $\Delta$ in the conduction band [see, e.g., J. P. Echeverry, B. Urbaszek, T. Amand, X. Marie, and I. C. Gerber, arXiv:1601.07351].






\end{widetext}


\begin{thebibliography}{99}
\bibitem{footnote_thanks}  Details on the fabrication and measurement of the gated samples are found in Refs.~\cite{Jones_NatNano13} and ~\cite{Ross_NatComm13}.
\bibitem{footnote1} For the bare potential, I have used the conventional 2D form, $V_{\mathbf{q}} = 2\pi e^2 / A \epsilon_r q$, where $A$ is the quantization area. Note that the geometrical correction to the dielectric screening introduced by Keldysh \cite{Keldysh_JETP79} is relevant when $qd \ll 1$ ($d \sim 1$~nm is the monolayer thickness including the van der Waals gap).  This condition breaks for $q=K_0$, at which Eq. (1) in Ref.~\cite{Keldysh_JETP79} conforms to the conventional 2D form. As for intravalley Coulomb interaction in the limit $q \rightarrow 0$, the dominant contribution comes from the inverse dependence on the Thomas-Fermi screening wavenumber at elevated densities.
\bibitem{supp} At the end of this document, I added notes on the derivation of the matrix element in (\ref{eq:orbital_mat_ele}), the plasmon dispersion in (\ref{eq:plasmon_modes})  and the region of free-plasmon propagation,  and the conductivity sum rule that leads to (\ref{eq:SPP}). I also added details on the calculation of the band-gap renormalization ($\Sigma_0$ in (\ref{eq:delta_comp}) and Fig.~2(d)), and provided a longer discussion on the relevance of intervalley plasmons in electron rather than hole rich systems. This material includes Refs.~\cite{Wang_PRB14}-\cite{Robert_arXiv16}. 
\bibitem{footnote2} The population of dark excitons can be neglected by assuming slow spin relaxation compared with the recombination lifetime. This assumption is justified given that the measured spin relaxation at low temperatures \cite{Yang_NatPhys15,Bushong_arxiv16} is about three order of magnitude longer that the measured recombination lifetime \cite{Wang_PRB14,Wang_APL15,Robert_arXiv16}. In addition, one difference from the elevated electron density case is that the redshift due to long-range excitations ($\Sigma_0$) should be reevaluated if the background electron density is smaller than the exciton density \cite{Haug_SchmittRink_PQE84}.
\end{thebibliography}
\end{document}